\documentstyle[12pt,epsfig,amsfonts]{article}
\newcommand{\beq}{\begin{eqnarray}}
\newcommand{\eeq}{\end{eqnarray}}
\newcommand{\eqn}{\begin{equation}}
\newcommand{\een}{\end{equation}}
\begin{document}
\title{Adiabatic motion of two-component BPS kinks}
\author{A. Alonso Izquierdo$^{(a)}$,
M.A. Gonzalez Leon$^{(a)}$, \\ J. Mateos Guilarte$^{(b)}$ and M.
de la Torre Mayado$^{(b)}$
\\ {\normalsize {\it $^{(a)}$ Departamento de Matematica
Aplicada}, {\it Universidad de Salamanca, SPAIN}}\\{\normalsize
{\it $^{(b)}$ Departamento de Fisica}, {\it Universidad de
Salamanca, SPAIN}}}

\date{}
\maketitle
\begin{abstract}
The low energy dynamics of degenerated BPS domain walls arising in
a generalized Wess-Zumino model is described as geodesic motion in
the space of these topological walls.
\end{abstract}

\section{Introduction}
According to a fruitful idea by Manton, geodesics in the moduli
space determine the slow motion of topological defects ,
\cite{Manton}. The adiabatic principle \cite{Uhlenbeck} has been
successfully applied to black holes \cite{Gibbons}, magnetic
monopoles \cite{Atiyah}, and self-dual vortices, both in Higgs,
\cite{Samols}, and Chern-Simons-Higgs,
\cite{Dziarmaga}-\cite{GG}-\cite{Manton2}-\cite{Kim}, models.
Recently, the moduli space of BPS domain walls has been discussed
by Tong \cite{Tong} and, following Manton's method, the low energy
dynamics of solitons has been studied by Townsend and Portugues
\cite{Townsend} in variations of the Wess-Zumino model.

In this paper we shall study the low energy dynamics of BPS kinks
living in a topological sector of a super-symmetric
(1+1)-dimensional system proposed by Bazeia and co-workers in
\cite{Bazeia}. In \cite{Voloshin} Shifman and Voloshin  have shown
that the (1+1)D system comes from the dimensional reduction of a
generalized Wess-Zumino model with two chiral super-fields; in
this latter case, the kink solutions appear as BPS domain walls.

From a one-dimensional perspective, the system encompasses several
topological sectors and Shifman and Voloshin in \cite{Voloshin}
discovered that there exists a degenerate family of BPS domain
walls in a distinguished sector. In \cite{Aai0}, three of us found
that for some critical values of the coupling parameter there
exist more degenerate kink families in other topological sectors,
although the new walls are generically non-BPS.

Each BPS domain wall, however, seems to be made from two basic
walls, which belong to other topological sectors. This interesting
structure has been explored by Sakai and Sugisaka, who found in
\cite{Sakai} an intriguing bound-state of wall/anti-wall pairs.
The aim of this work is to describe how the basic kinks move in
the space of BPS topological kinks. To achieve this goal we shall
apply Manton's method and we shall thus extend the applicability
of the adiabatic principle to one-dimensional topological defects.

The organization of the paper is as follows: in Section \S 2 we
briefly summarize the general framework of (1+1)D super-symmetric
field theory. Section \S 3 is devoted to describing the moduli
space of BPS super-kink solutions. In Section \S 4 we unveil the
metric inherited from the adiabatic kink motion in the space of
kink solutions, determine the geodesic orbits, and describe the
low speed motion of BPS kinks. Finally, in Section \S 5 we
interpret these results from the point of view of moving walls.

\section{${\cal N}=1$ super-symmetric (1+1)-dimensional field-theoretical systems}
Using non-dimensional space-time coordinates and fields as defined
in Reference \cite{Aai0}, we consider Bose,
$\hat{\vec{\phi}}(x^\mu )=\sum_{a=1}^2\hat{\phi}^a(x^\mu
)\vec{e}_a$, and Fermi,
\[
\hat{\vec{\psi}}(x^\mu)=\sum_{a=1}^2\left(\begin{array}{c}\hat{\psi}_1^a(x^\mu
) \\\hat{\psi}_2^a(x^\mu )\end{array}\right)\vec{e}_a \qquad ,
\]
fields. Here, $x^\mu=(x^0,x^1)$ are local coordinates in ${\Bbb
R}^{1,1}$ space-time; $\vec{e}_a.\vec{e}_b=\delta_{ab}$ is an
ortho-normal basis in ${\Bbb R}^2$ internal (iso-spin) space, and
the Fermi fields belong to the Majorana representation of the
${\rm Spin}(1,1;{\Bbb R})$ group: if $\{\gamma^\mu ,\gamma^\nu \}=
2g^{\mu\nu}$, $g^{\mu\nu}={\rm diag}(1, -1)$, is the Clifford
algebra of ${\Bbb R}^{1,1}$, we choose $\gamma^0=\sigma^2 ,
\gamma^1=i\sigma^1$. Also, $\gamma^5=\gamma^0\gamma^1=\sigma^3$,
with $\sigma^1,\sigma^2,\sigma^3$ the $2\times 2$ Pauli matrices.

The canonical quantization procedure dictates the equal-time
commutation/anticommutation relations among the fields and their
momenta. Defining $\hat{\vec{\pi}}(x^\mu
)=\frac{\partial\hat{\vec{\phi}}}{\partial x^0}(x^\mu )$, we have
that:
\begin{equation}
[\hat{\phi}^a(x),\hat{\pi}^b(y)]=i\delta^{ab}\delta(x-y)\qquad ,
\qquad
\{\hat{\psi}^a_\alpha(x),\hat{\psi}^b_\beta(y)\}=\delta^{ab}\delta_{\alpha\beta}\delta(x-y)
\quad , \label{eq:quan}
\end{equation}
where $\alpha,\beta=1,2$ are Majorana spinor indices and a natural
system of units -$\hbar=c=1$- has been chosen.

Interacting ${\cal N}=1$ super-symmetric field theory is built
from the normal-ordered super-charge operator:
\begin{equation}
\hat{Q}=\int dx  \,\colon
\left[\gamma^\mu\gamma^0\hat{\vec{\psi}}(x^\mu
)\partial_\mu\hat{\vec{\psi}}(x^\mu
)+i\gamma^0\hat{\vec{\psi}}(x^\mu
)\vec{\nabla}\hat{W}(x^\mu)\right]\colon \qquad .\label{eq:sup}
\end{equation}
Interactions come from the gradient of the super-potential
$\vec{\nabla}\hat{W}(x^\mu)=\sum_{a=1}^2\hat{\frac{\partial
W}{\partial \phi^a}}(x^\mu )\vec{e}_a$ and the super-symmetry
algebra:
\begin{equation}
\{\hat{Q}_\alpha,\hat{Q}_\beta\}=2(\gamma^\mu\gamma^0)_{\alpha\beta}\hat{P}_\mu
-2i\gamma^1_{\alpha\beta}\hat{T}\quad , \quad \alpha ,\beta=1,2
\label{eq:algs}
\end{equation}
encompasses the energy,
\begin{equation}
\hat{P}_0=\frac{1}{2}\int dx
\,\colon\left[\hat{\vec{\pi}}\hat{\vec{\pi}}+\frac{\partial\hat{\vec{\phi}}}{\partial
x}\frac{\partial\hat{\vec{\phi}}}{\partial
x}+\vec{\nabla}\hat{W}\vec{\nabla}\hat{W}-i\hat{\vec{\bar{\psi}}}\gamma^1\frac{\partial
\hat{\vec{\psi}}}{\partial
x}+\hat{\vec{\bar{\psi}}}\vec{\vec{\Delta}}\hat{W}\hat{\vec{\psi}}\right]\colon
\qquad , \label{eq:ener}
\end{equation}
the momentum,
\begin{equation}
\hat{P}_1=-\int
dx\,\colon\left[\hat{\vec{\pi}}\frac{\partial\hat{\vec{\phi}}}{\partial
x}+\frac{i}{2}\hat{\vec{\psi^t}}\frac{\partial\hat{\vec{\psi}}}{\partial
x} \right]\colon \qquad ,\label{eq:mom}
\end{equation}
and the anomalous topological/central charge,
\begin{equation}
\hat{T}=\int dx\,
\colon\left[\frac{\partial\hat{\vec{\phi}}}{\partial
x}.\vec{\nabla}\left(\hat{W}+\frac{1}{4\pi}\Delta\hat{W}\right)\right]\colon
\quad , \label{eq:top}
\end{equation}
operators. In formulas
(\ref{eq:ener})-(\ref{eq:mom})-(\ref{eq:top}) we have defined
$\hat{\vec{\bar{\psi}}}(x^\mu )=\hat{\vec{\psi^t}}(x^\mu
)\gamma^0$, and
\[
\vec{\vec{\Delta}}\hat{W}=\vec{\nabla}\otimes\vec{\nabla}\hat{W}=\sum_{a=1}^2\sum_{b=1}^2
\vec{e}_a\otimes\vec{e}_b\frac{\partial^2\hat{W}}{\partial\phi^a\partial\phi^b}\quad
,
\Delta\hat{W}=\vec{\nabla}\vec{\nabla}\hat{W}=\sum_{a=1}^2\frac{\partial^2\hat{W}}
{\partial\phi^a\partial\phi^a}
\]
are respectively the Hessian and Laplacian operators applied to
$\hat{W}$. Note that there is no anomaly in the central charge if
the super-potential is harmonic, the condition for the existence
of ${\cal N}=2$ super-symmetry.

The usual definition of BPS states in this system, \cite{Olive},
is derived from (\ref{eq:algs}):
\begin{equation}
\hat{P}_0=\frac{1}{2}\left(\hat{Q}_1\pm\hat{Q}_2\right)^2+|\hat{T}|\qquad
, \qquad \left(\hat{Q}_1\pm\hat{Q}_2\right)|BPS\rangle =0
\label{eq:bps}
\end{equation}
as the requirement of minimal energy in each super-selection
sector. To find these states a variational method, using the
coherent states
\[
\hat{\vec{\pi}}(x)|\vec{\phi}(x),\vec{\psi}(x)\rangle =\vec{0}\,\,
, \, \hat{\vec{\phi}}(x)|\vec{\phi}(x),\vec{\psi}(x)\rangle
=\vec{\phi}(x)|\vec{\phi}(x),\vec{\psi}(x)\rangle \,\, , \,
\hat{\vec{\psi}}(x)|\vec{\phi}(x),\vec{\psi}(x)\rangle
=\vec{\psi}(x)|\vec{\phi}(x),\vec{\psi}(x)\rangle
\]
as trial states, is conventionally applied. $\vec{\phi}(x)$ and
$\vec{\psi}(x)$ are respectively scalar and Majorana spinor static
classical field configurations and, on these states, the
(\ref{eq:bps}) BPS condition becomes:
\begin{equation}
\vec{\psi}_\mp (x)\left( \frac{d\vec{\phi}}{dx}\pm \vec{\nabla}W
\right)|\vec{\phi}(x),\vec{\psi}(x)\rangle =0 \qquad , \,
\vec{\psi}_\pm(x)=\vec{\psi}_1(x)\pm \vec{\psi}_2(x)
\label{eq:cbps1}
\end{equation}
because the expectation values in this kind of state of
normal-ordered operator functionals are equal to their classical
counterparts, see e.g. \cite{Coleman}.

Thus, the BPS states are the coherent states for which the scalar
field configurations correspond to the flow lines of $\mp{\rm
grad}W$, whereas the corresponding $\gamma^1$ eigen-spinor
$\vec{\psi}_\pm$ configurations are zero. Moreover, super-symmetry
forces the surviving (non-null) spinor configurations to satisfy
the equation:
\begin{equation}
\frac{d\vec{\psi}_\mp}{dx}(x)=-\vec{\vec{\Delta}}W(\vec{\phi})\vec{\psi}_\mp
(x)\label{eq:cbps2}
\end{equation}
\section{The BPS kink moduli space}

Our choice of super-potential is:
\[
W(\vec{\phi})=4\sqrt{2}\left[\frac{1}{3}(\vec{\phi}\cdot
\vec{e}_1)^3-\frac{1}{4} \vec{\phi}\cdot
\vec{e}_1(1-(\vec{\phi}\cdot\vec{e}_2)^2)\right] \qquad .
\]
Thus, we deal with the super-symmetric extension of the BNRT model
analyzed in \cite{Aai0} for the special value of the parameter
$\sigma=\frac{1}{2}$.

In Reference \cite{Aai0} the BPS kinks were shown to be in
one-to-one correspondence with the kink orbits:
\begin{equation}
\phi_1^2 + \frac{1}{2} \phi_2^2 = \frac{1}{4}+ c \phi_2^4 \qquad .
\label{eq:traba1}
\end{equation}
The explicit analytical expressions for these families of BPS
kink/anti-kink solutions are:
\begin{equation}
\vec{\phi}^K [x;a,b]=\pm\left(\frac{1}{2} \frac{\sinh\left(
2\sqrt{2} (x+a)\right)}{\cosh\left( 2
\sqrt{2}(x+a)\right)+b^2}\vec{e}_1+ \frac{b}{\sqrt{b^2+\cosh\left(
2 \sqrt{2}(x+a)\right)}}\vec{e}_2\right) \label{eq:bbpsk}
\end{equation}
for arbitrary integration constants $a,b \in (-\infty ,\infty )$.
The constant $b$ is defined in terms of $c\in (-\infty
,\frac{1}{4})$ as: $b=\pm\sqrt{\frac{1}{\sqrt{1-4c}}} $, see
\cite{Aai0}. Thus, the plane is the space ${\cal S}^K$ of BPS
kinks : there is a one-to-one correspondence between the points
$(a,b)\in{\cal S}^K\simeq{\Bbb R}^2$ and the space of topological
kink solutions of the field equations. The $\phi_1$- and
$\phi_2$-components of these solitary waves are shown in Figure 1
for $a=0$ and several particular values of $b$. The dependence on
$a$ of the kink profile is trivial; the value of $a$ only
determines the \lq\lq center" of the kink.

\noindent
\begin{figure}[htbp]
\centerline{ \epsfig{file=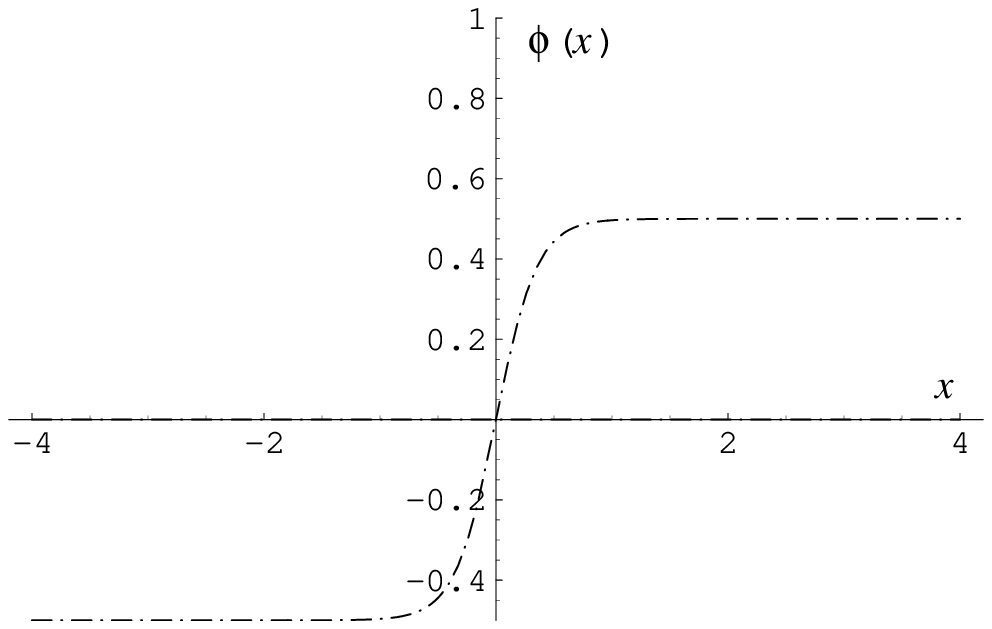, height=2.5cm}
\epsfig{file=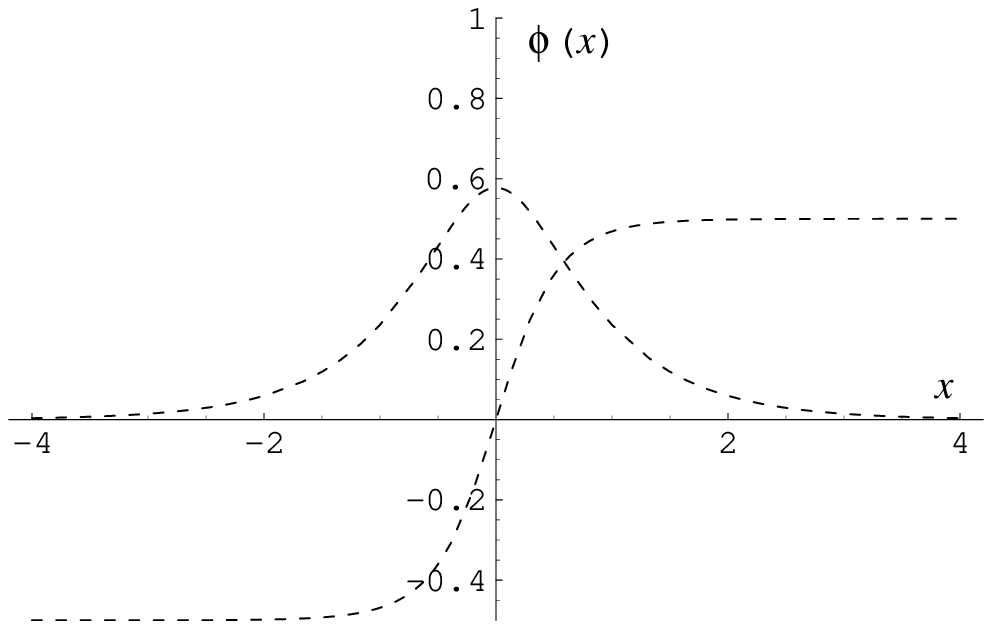, height=2.5cm} \epsfig{file=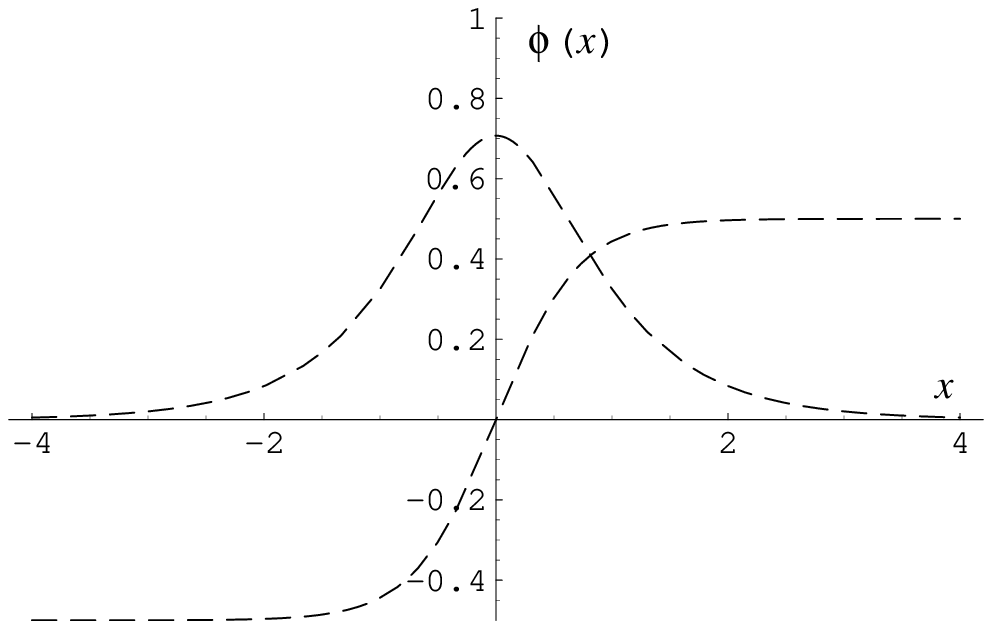,
height=2.5cm} \epsfig{file=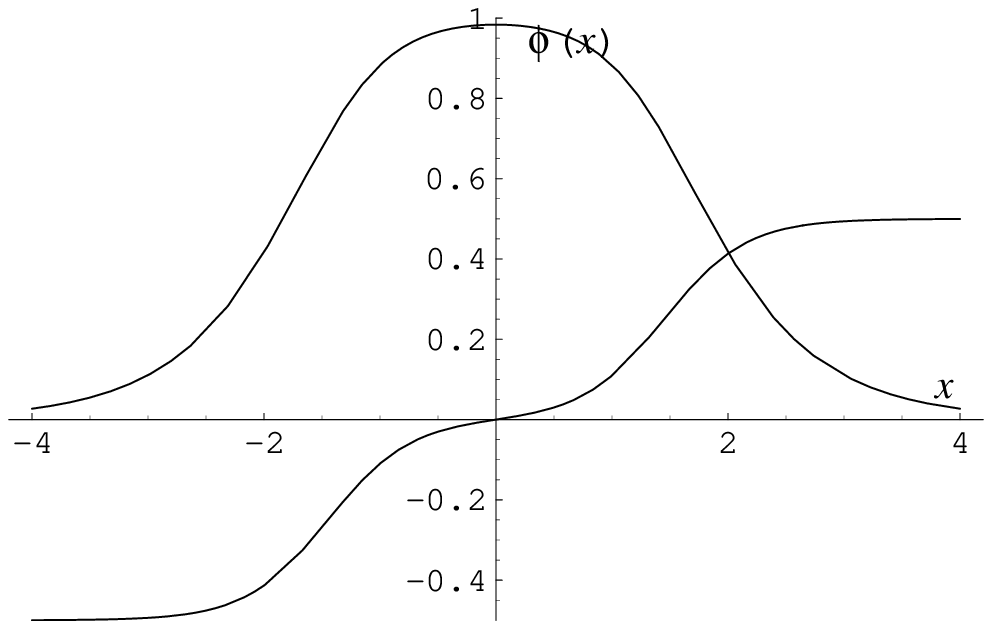, height=2.5cm}}
\caption{\small Solitary waves (\ref{eq:bbpsk}) corresponding to:
{\it (a)} $b=0$, {\it (b)} $b=\sqrt{0.5}$, {\it (c)} $b=1$ and
{\it (d)} $b=\sqrt{30}$.}
\end{figure}
This sequence of Figures shows us the main characteristics of the
different BPS kinks. If $b=0$ only one of the two field components
-$\phi_1^K$- is different from zero and the kink configuration
interpolates between the minimum, $A_+=\frac{1}{2}\vec{e}_1$, and
the maximum, $A_-=-\frac{1}{2}\vec{e}_1$, of $W$, the vacua of the
theory. Therefore, the $(a,0)$ kinks are topological solutions
with one non-null component and, for obvious reasons,  in the
literature they are called TK1 kinks. If $b\neq 0$, $\phi_2^K$ is
also different from zero but the field configurations still
interpolate between the $A_+=\frac{1}{2}\vec{e}_1$ and
$A_-=-\frac{1}{2}\vec{e}_1$ vacua. All these topological kinks
have the two field components different from zero and they are
thus called TK2 kinks. Note that changing $b$ to $-b$ merely
amounts to changing $\phi_2^K$ to $-\phi_2^K$.

The associated BPS fermionic form factors, the solutions of
(\ref{eq:cbps2}) for $\vec{\phi}^K$, are:
\begin{equation}
\vec{\psi}^K_{\pm}(x)=0 \qquad , \quad \vec{\psi}^K_{\mp}[x;a,b ;
d,f]=\left[ d\frac{\partial\vec{\phi}^K}{\partial
a}[x;a,b]+f\frac{\partial\vec{\phi}^K}{\partial
b}[x;a,b]\right]\varepsilon_{\mp}\quad , \label{eq:fbpsk}
\end{equation}
where $d,f$ are real integration constants and
$\varepsilon_{\mp}=\varepsilon^{K^{\mp}}_1\mp\varepsilon^{K^{\mp}}_2=2$;
$\varepsilon^{K^{\mp}}=\left(\begin{array}{c}1\\ \mp
1\end{array}\right)$ are constant eigen-spinors of $\gamma^1$. (
See Figures 2 and 3, where the fermionic partners of the bosonic
solitary waves for the same values of the $b$ parameter as above,
are plotted ) .

\noindent
\begin{figure}[htbp]
\centerline{ \epsfig{file=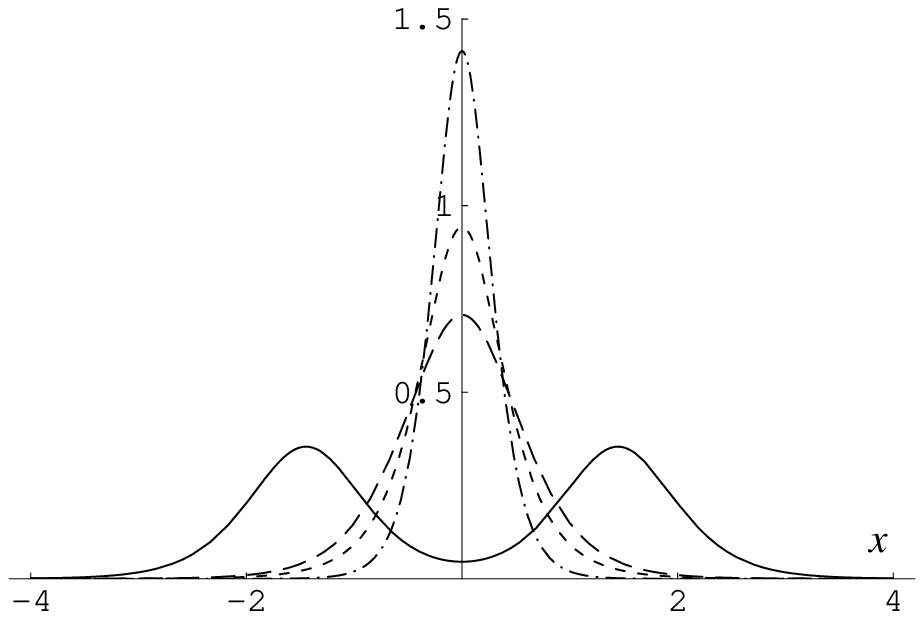, height=3.5cm}
\epsfig{file=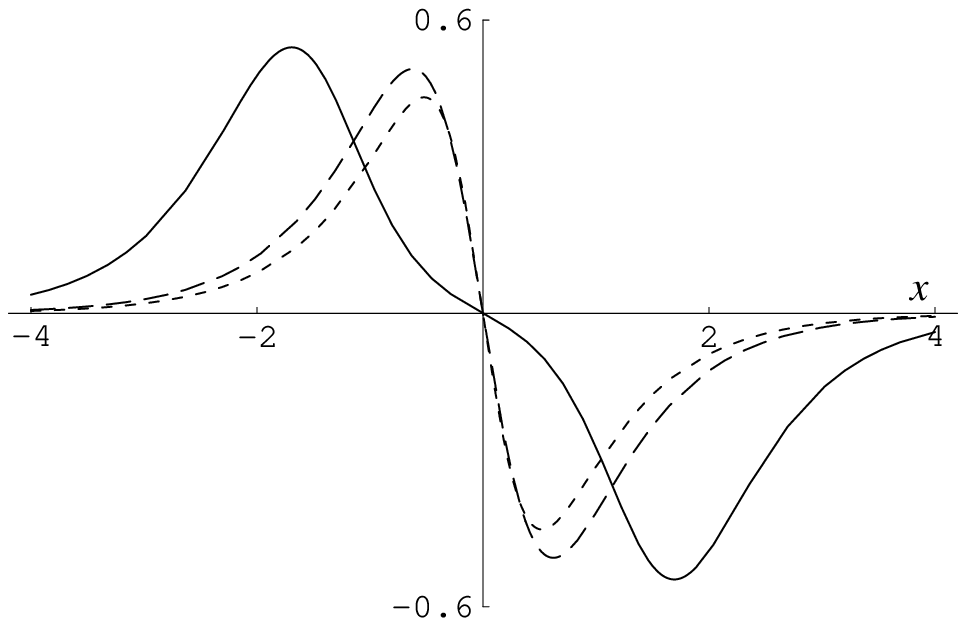, height=3.5cm}}\caption{\small First {\it
(a)} and second {\it (b)} components of $\frac{\partial
\vec{\phi}^K}{\partial a}$.}
\end{figure}

\noindent
\begin{figure}[htbp]
\centerline{ \epsfig{file=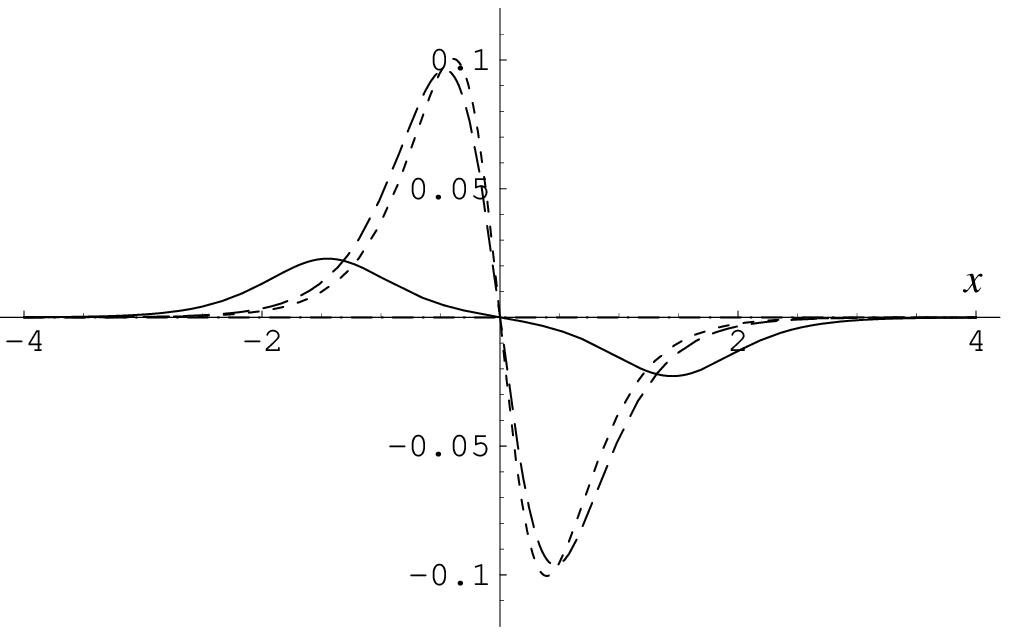, height=3.5cm}
\epsfig{file=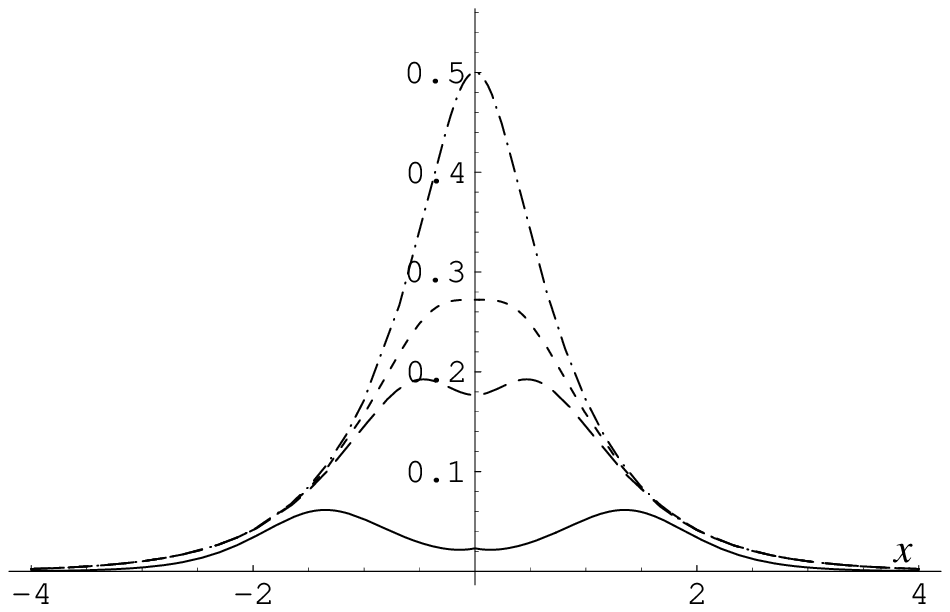, height=3.5cm} } \caption{\small First {\it
(a)} and second {\it (b)} components of $\frac{\partial
\vec{\phi}^K}{\partial b}$.}
\end{figure}

The Majorana-Weyl spinors shown in (\ref{eq:fbpsk}) span the
tangent space to the space of BPS kink solutions $T{\cal
S}^K\simeq T{\Bbb R}^2$ because they are linear combinations of
the bosonic zero modes. Moreover, these fermionic configurations
also solve the static Dirac equation of the coupled system and
therefore, do not contribute to $\langle
\vec{\psi}(x),\vec{\phi}(x)|\colon \hat{P}_0\colon
|\vec{\phi}(x),\vec{\psi}(x)\rangle$; the BPS super-kinks, formed
by the combination of the bosonic and fermionic solutions, also
saturate the topological bound.

The properties of a solitary wave determined by a point in ${\cal
S}^K$ are encoded in the bosonic energy density:
\begin{equation}
{\cal E}^K[x;a,b]=\frac{\partial\vec{\phi}^K}{\partial
x}.\frac{\partial\vec{\phi}^K}{\partial x}=\frac{4+7b^2{\rm
cosh}[2\sqrt{2}(x+a)]+4b^4{\rm cosh}[4\sqrt{2}(x+a)]+b^2{\rm
cosh}[6\sqrt{2}(x+a)]}{2(b^2+{\rm
cosh}[2\sqrt{2}(x+a)])^4}\label{eq:dens}
\end{equation}
Figure 4 shows a plot of ${\cal E}^K[x;a,b]$ for $a=0$ and the
same values of $b$ as in the above Figures. The identity ${\cal
E}^K[x;a,b]={\cal E}^K[x;a,-b]$ is a consequence of the invariance
of the theory under the action of the group $G={\Bbb Z}_2\times
{\Bbb Z}_2$ generated by the reflections $\phi_1\rightarrow
-\phi_1 , \phi_2\rightarrow -\phi_2$. $\phi_1\rightarrow -\phi_1$
merely replaces kinks by antikinks but taking a quotient by
$\phi_2\rightarrow -\phi_2$ both in (\ref{eq:traba1}) and
(\ref{eq:bbpsk}) we find the moduli space ${\cal M}^K\simeq{\cal
S}^K/{\Bbb Z}_2$ of BPS kinks. Thus, $a$ and $b^2$ are good
coordinates in ${\cal M}^K$, which is isomorphic to the upper
half-plane: ${\Bbb H}=(-\infty ,\infty)\times [0,\infty )$.

There are two regimes in the $b^2$-parameter classified by the
dependence on $b^2$ of the critical points of the energy density,
i.e. the zeroes of:
\begin{eqnarray}
\frac{\partial{\cal E}^K}{\partial x}[x;a,b]=2\left(
\frac{\partial\phi_1^K}{\partial
x}\frac{\partial^2\phi_1^K}{\partial^2x}\right. &+&\left.
\frac{\partial \phi_2^K}{\partial
x}\frac{\partial^2\phi_2^K}{\partial^2x}\right)=\frac{4\sqrt{2}
\sinh 2\sqrt{2}(x+a)}{(b^2+\cosh 2\sqrt{2}(x+a))^5}\ P_3(\cosh
2\sqrt{2}(x+a))\qquad \label{eq:max}\\ P_3(\cosh 2
\sqrt{2}(x+a))&=& -b^2\cosh^32\sqrt{2}(x+a)-b^4 \cosh^2
2\sqrt{2}(x+a)+\nonumber \\&+&(-3b^2+4b^6) \cosh
2\sqrt{2}(x+a)+5b^4-4 \qquad \nonumber .
\end{eqnarray}
Note that (\ref{eq:max}) relates the shape of the energy density
to the shape of the kink profile.

\noindent
\begin{figure}[htbp]
\centerline{ \epsfig{file=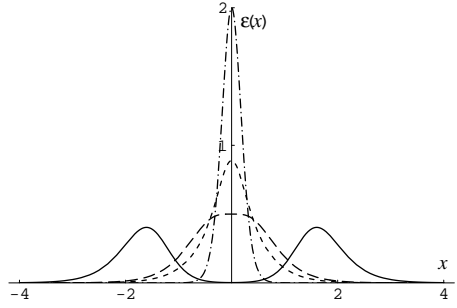, height=4cm} } \caption{\small
Energy density ${\cal E}^K[x;0,b]$.}
\end{figure}

Apart from the obvious solution, $x=-a$, i.e. $\sinh
2\sqrt{2}(x+a)=0$, $\forall b^2$, we can classify the solutions of
$\frac{\partial{\cal E}^K}{\partial x}[x;a,b]=0$ in terms of the
roots of the cubic polynomial $P_3(\cosh 2\sqrt{2}(x+a))$. Writing
$P_3$ as $P_3(\cosh 2\sqrt{2}(x+a))=-b^2\tilde{P}(u)$, where
$\tilde{P}(u)$ is the bi-cubic polynomial
$\tilde{P}(u)=(u^2)^3+(b^2+3) (u^2)^2-(4b^4-2b^2-6) (u^2)-4\left(
b^4+b^2-1-\frac{1}{b^2}\right)$ in the variable $u^2(x)=-1+\cosh
2\sqrt{2}(x+a)$, a classical analysis of the roots of
$\tilde{P}(u)$, based on the Cardano and the Vieta formulae and
use of Rolle's theorem, shows that:

\begin{itemize}
\item $\tilde{P}(u)$ has no real roots in $u(x)$ if $b^2\in [0,1]$.
Thus,
 $\frac{\partial{\cal E}^K}{\partial
x}[x;a,b]=0$ only for $x=-a$, which is the only critical point of
${\cal E}^K$ as a function of $x$. Moreover, ${\cal
E}^K[-a;a,b]=\frac{2}{(b^2+1)^2}$ is the maximum value of ${\cal
E}^K$ on the real line; $b^2$ therefore measures the height of the
solitary wave energy density , see Figure 4.

\item  Things are more interesting if $b^2\in (1,\infty)$: $\tilde{P}$ has two real
roots. As a cubic polynomial in $u^2$, $\tilde{P}$ has a single
positive root $r(b^2)$ that depends on the value of $b^2$; hence,
$u=\pm \sqrt{r}$ are the real roots of $\tilde{P}(u)$. Besides
$x=-a$, which is a minimum, two other critical points of ${\cal
E}^K$ arise at $x=-a\pm m(b^2)$ if $b^2>1$, where
$m(b^2)=\frac{1}{2\sqrt{2}} {\rm arccosh} (1+r(b^2))$ .

These two points are maxima of ${\cal E}^K$ and the solitary wave
is made from two lumps if $b^2>1$. The distance between the peaks
grows with $b^2$, and $b^2$ must be understood as the relative
coordinate of a system of two \lq\lq particles" if $b^2>1$,
whereas $-a$ is still the center of mass coordinate.

\end{itemize}

Alternatively, one can trust Mathematica and just look at Figure
4.

Exactly at the $b=\pm\infty$ limits, the solutions
\begin{equation}
\vec{\phi}^{B_\pm}[x;x_0, \alpha , \beta ]=(-1)^\alpha \left(
\frac{1}{4}(1-(-1)^\beta {\rm tanh}(\sqrt{2}(x+x_0))\vec{e}_1\pm
\sqrt{\frac{1}{2}(1+(-1)^\beta{\rm
tanh}(\sqrt{2}(x+x_0))}\vec{e}_2\right)\ , \label{eq:tkab}
\end{equation}
($\alpha,\beta,=0,1$) living in other topological sectors, appear,
see Figure 5. The integration constant $x_0\in{\Bbb R}$ determines
the center of mass of the lump: $x_{{\rm
CM}}=-x_0+\frac{1}{2\sqrt{2}} \ln \frac{1+\sqrt{17}}{8}$ -the
shift with respect to what one expects for the center of mass is
seen in Figure 5(b). Note that the inflection points in $\phi_1^B$
and $\phi_2^B$ do not coincide. $\vec{\phi}^{B_\pm}[x;x_0, \alpha,
\beta]$ are also flow lines of ${\rm grad}W$ but start or end at a
saddle point $\pm\vec{e}_2$ of $W$ -the other two vacua of the
theory- . \noindent
\begin{figure}[htbp]
\centerline{ \epsfig{file=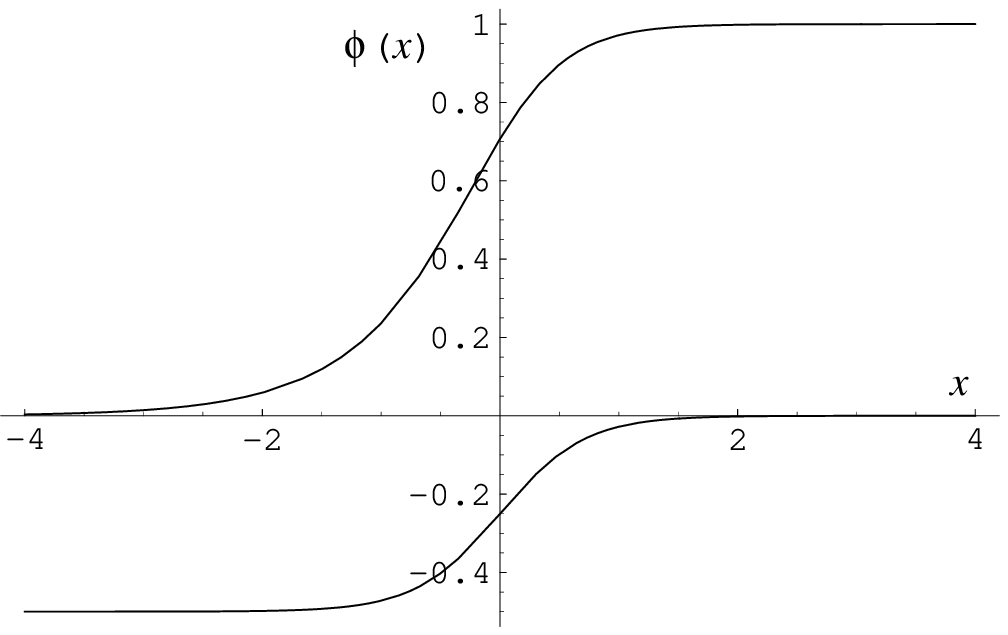,
height=3.5cm}\epsfig{file=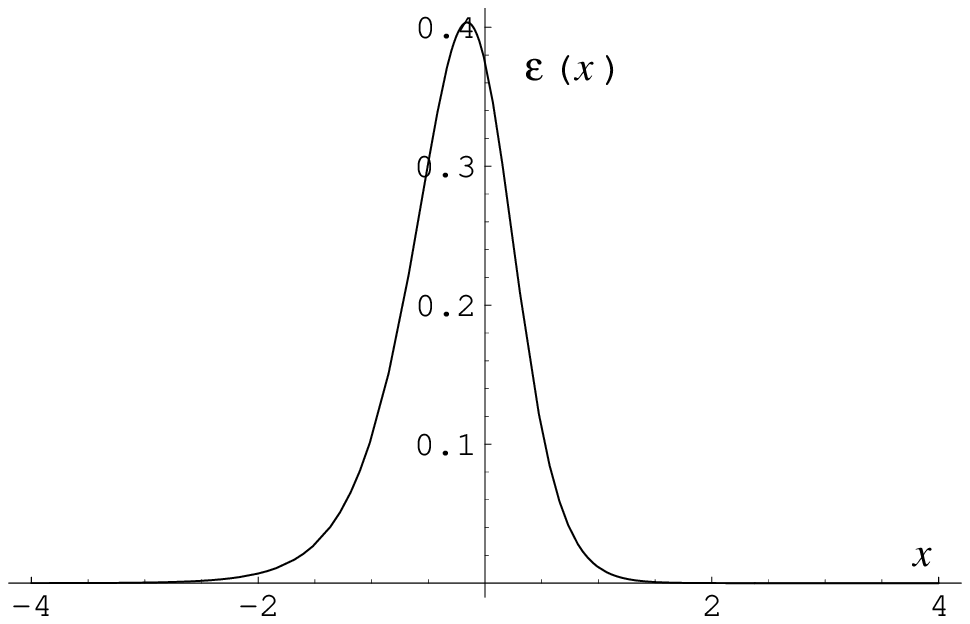, height=3.5cm}}
\caption{\small One of the solitary waves (\ref{eq:tkab}) {\it
(a)} and its energy density {\it (b)}.}
\end{figure}
The shape of the energy density for large values of $|b|$ suggests
that the $\vec{\phi}^{B_\pm}$ solitary waves are the basic kinks
in the topological sector of the $\vec{\phi}^K$ family. Near
$|b|=\infty$ the identity
\begin{equation}
\vec{\phi}^K[x;a,b]{\simeq\atop
b\rightarrow\pm\infty}\vec{\phi}^{B_\pm}[x;x_0^+,1,0]\,
\theta(-x-a)+ \vec{\phi}^{B_\pm}[x;x_0^-,0,1]\, \theta(x-a)\quad ,
\label{eq:mei}
\end{equation}
\[
x_0^\pm =-a\pm m(b^2)+\frac{1}{2\sqrt{2}}\ln \frac{1+\sqrt{17}}{8}
\]
where $\theta(z)$ is the Heaviside step function, approximately
holds. In fact, one can prove that the approximation becomes exact
at the $|b|=\infty$ limit, where the two $\vec{\phi}^{B_\pm}$
kinks become infinitely separated. Thus, they are not accessible
in the kink space ${\cal S}^K$ but belong to the boundary; the
circle of infinite radius in ${\Bbb R}^2$: $\partial{\cal
S}^K=S^1_\infty$.

Besides $b=0$ and $b=\pm\infty$, there are two other special
points in ${\cal S}^K$: $b=\pm 1$. The corresponding kink orbits
are the upper and lower half-ellipses
$\phi_1^2+\frac{1}{2}\phi_2^2=\frac{1}{4}$ and we shall use the
term  TK2E$_\pm$ to refer to these two-component kinks, which form
the frontier in ${\cal M}^K$ between solitary waves carrying one
or two lumps of energy density.

\section{Geodesic motion in the kink space $
{\cal S}^K$} In our framework, the adiabatic principle is
equivalent to restricting time-evolution to the sub-space of BPS
kink states. The expectation value of the kinetic energy in these
states -there is no contribution of the fermionic variables to the
kinetic energy because the Dirac Lagrangian is first-order in time
derivatives- is:
\begin{eqnarray}
K=\frac{1}{2}\int &dx& \langle \vec{\phi}^K[x;a(t),b(t)]| \colon
\hat{\vec{\pi}}(x)\hat{\vec{\pi}}(x)\colon |
\vec{\phi}^K[x;a(t),b(t)]\rangle
=\nonumber\\&=&\frac{1}{2}g_{aa}(a,b)\dot{a}^2+g_{ab}(a,b)\dot{a}\dot{b}+
\frac{1}{2}g_{bb}(a,b)\dot{b}^2 \label{eq:adiabm}
\end{eqnarray}
where:
\begin{equation}
g_{aa}(a,b)=\int_{-\infty}^{\infty} dx \frac{\partial
\vec{\phi}^K}{\partial a}\cdot \frac{\partial
\vec{\phi}^K}{\partial a};\quad g_{ab}(a,b)=
\int_{-\infty}^{\infty} dx \frac{\partial \vec{\phi}^K}{\partial
a}\cdot \frac{\partial \vec{\phi}^K}{\partial b};\quad
g_{bb}(a,b)=\int_{-\infty}^{\infty} dx \frac{\partial
\vec{\phi}^K}{\partial b}\cdot \frac{\partial
\vec{\phi}^K}{\partial b}\label{eq:int}
\end{equation}

We think of $K$ as the Lagrangian for geodesic motion in the kink
space ${\cal S}^K$ with a metric inherited from the dynamics of
the bosonic zero modes. Alternatively, the fermionic partners of
the BPS kinks acquire a direct geometrical meaning; they are the
zweig-beins ( \lq\lq the square root" ) of the metric  in the
sense that,
\begin{eqnarray*}
g_{11}(a,b)&=&\int dx \, \vec{\psi}_{\mp}^K[x; a,b ; 1,0]
\vec{\psi}_{\mp}^K[x; a,b ; 1,0]\quad ; g_{22}(a,b)=\int dx
\,\vec{\psi}_{\mp}^K[x; a,b ; 0,1] \vec{\psi}_{\mp}^K[x; a,b ;
0,1] \\ g_{12}(a,b)&=&\int dx \vec{\psi}_{\mp}^K[x; a,b ; 1,0]
\vec{\psi}_{\mp}^K[x; a,b ; 0,1]=g_{21}(a,b) \qquad .
\end{eqnarray*}
The metric tensor is ${\Bbb Z}_2$-symmetric - invariant under
$b\rightarrow -b$ -:
\begin{eqnarray}
g_{aa}(a,b)&=&\frac{2\sqrt{2}}{3}\quad ;\quad g_{ab}(a,b)=0 \quad
; \quad
g_{bb}(a,b)=\frac{2\sqrt{2}}{3}h(b)\nonumber\\h(b)&=&\frac{1}{4(b^4-1)^2}\left[
2 b^6-5b^2+3\frac{\arctan \left(
\frac{\sqrt{1-b^4}}{b^2}\right)}{\sqrt{1-b^4}}\right]
\label{eq:metric}\quad .
\end{eqnarray}
Computation of the integrals (\ref{eq:int}) has been performed by
changing variables to $y={\rm exp}[2\sqrt{2}(x+a)]$ in such a way
that the quadratures reduce to rational definite integrals in $y$.
The transition from one to two lumps is seen in formula
(\ref{eq:metric}) in the trading of ${\rm arctan}$ by $i\, {\rm
arctanh}$ that happens at $b=\pm 1$. $h(b)$, however, is also real
for $|b|>1$ because the denominator becomes purely imaginary in
this regime, see Figure 6.

\noindent
\begin{figure}[htbp]
\centerline{ \epsfig{file=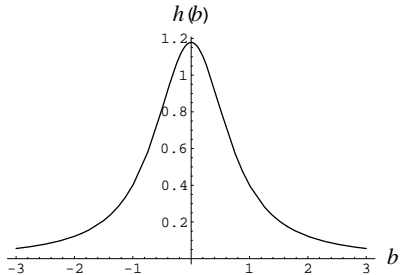, height=3.5cm}}\caption{\small
Graphic of the function $h(b)$.}
\end{figure}

There is an important difference with the low energy dynamics of
other topological defects. In this case the moduli space ${\cal
M}^K$ is an orbifold: the orbit of every point in the TK1 line
$(a,b=0)\in{\cal S}^K$ by the action of the ${\Bbb Z}_2$-group is
a single point whereas any other point in ${\cal S}^K$ is not
invariant under the $b\rightarrow -b$ reflection. Therefore,
geodesic motion in ${\cal M}^K$ would lead to non-smooth dynamics
at the TK1 line. Fortunately, the generic orbit is a two-element
set and one can safely pursue the analysis of geodesic motion in
${\cal S}^K$.

The geodesics of a metric in the kink space ${\cal S}^K$ of the
form :
\[
ds^2=\frac{2\sqrt{2}}{3}\left( da^2+h(b) \, db^2\right)
\]
are easily found ( at least implicitly). Writing the metric in
terms of a new variable, $\bar{b}=\int \sqrt{h(b)} db$, we have :
\[
\frac{3}{2\sqrt{2}} \, ds^2=da^2\, + \, d\bar{b}^2 \quad , \quad
d\bar{b}\, =\, \sqrt{h(b)} \, db
\]
and the geodesic curves are straight lines in the $a-\bar{b}$
plane. If $k_1$, $k_2$, $k_1'$, $k_2'$, are integration constants,
the geodesics are:
\begin{equation}
a(t)= k_1\, t\, +\, k_2 \qquad , \qquad  \bar{b}(t)= \int
\sqrt{h(b)} \, db=\, k_1' \, t+k_2'\quad .
\end{equation}
In terms of the new integration constants,
$\kappa_1=\frac{k_1'}{k_1}\, , \, \kappa_2=k_2'-\kappa_1k_2 \,$,
we can also represent the geodesic paths by writing $\bar{b}$ as a
function of $a$:
\begin{equation}
\bar{b}=\int \sqrt{h(b)}db=\kappa_1 a+\kappa_2 \qquad .
\label{eq:orbits}
\end{equation}
We immediately identify a simple kind of geodesic motion: taking
$\kappa_1=0$ in (\ref{eq:orbits}), orbits with $b={\rm constant}$
are found. This motion corresponds to free displacement of the
kink center of mass, for all types of solitary wave, without
changing the shape of the kink profile, see Figure 7.

\noindent
\begin{figure}[htbp]
\centerline{ \epsfig{file=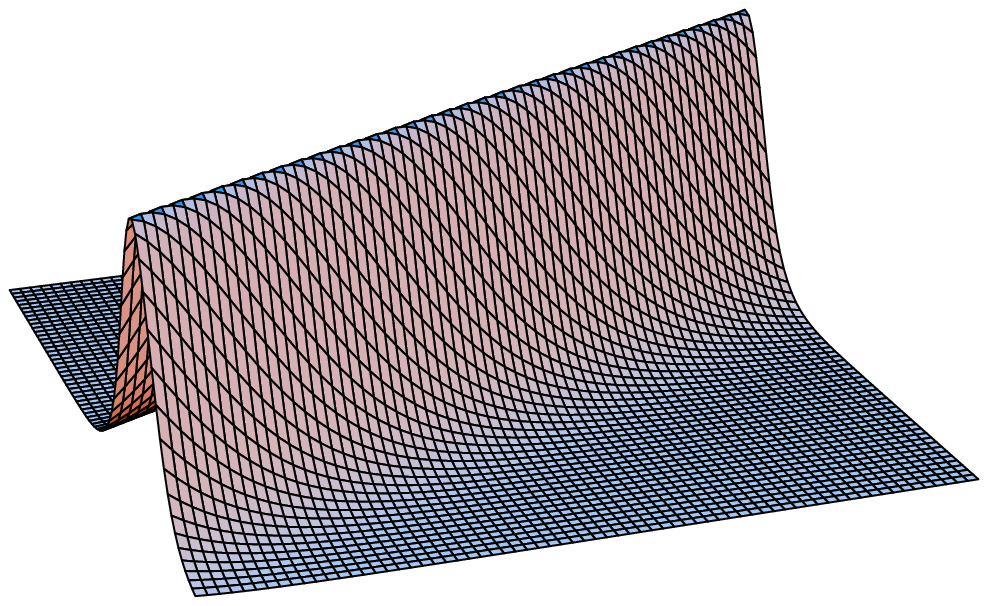, height=4cm}
\epsfig{file=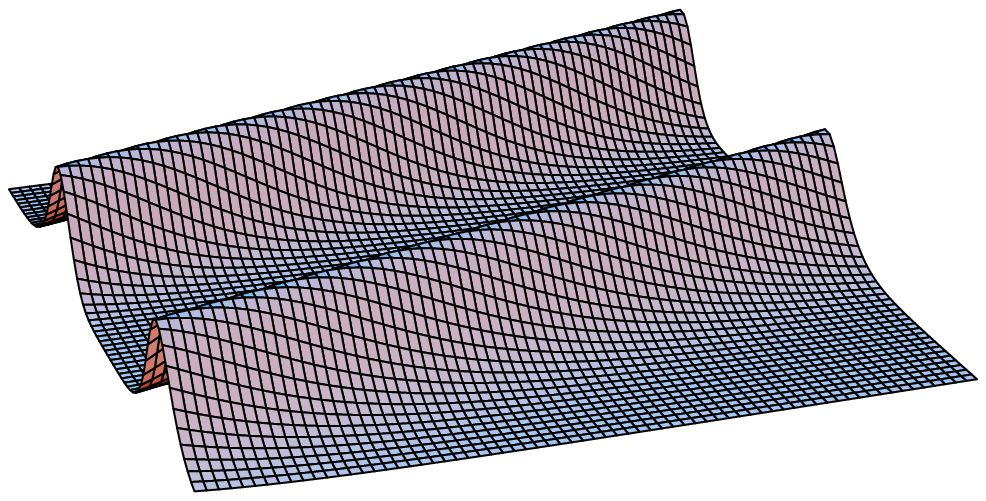, height=4cm} }\caption{\small Energy
density evolution along straight geodesic lines with $b$=constant:
{\it (a)} $b=0.9$, a single lump is moving {\it (b)} $b=10$,
synchronous motion of two lumps. Time runs from left to right }
\end{figure}

In a search for the more general geodesic motion, the problem is
to find the explicit form of $\int \sqrt{h(b)} db$. Because
performing this integration explicitly is out of reach by
analytical means, we propose two alternative ways for describing
the geodesic motion of BPS kinks.
\subsection{Numerical integration}
In Figure 8(a) a Mathematica numerical plot of several geodesic
orbits (\ref{eq:orbits}) is shown. $\kappa_1$ has been set to 3,\,
2 and 1, whereas the freedom in $\kappa_2$ has been fixed by
setting the value of $b=0.1$ -near the TK1 point in the space
${\cal S}^K$- at the instant $t=-\frac{k_2}{k_1}$. The features
common to these generic geodesics are as follows: the starting
point at $t=-\infty$ is the point in $\partial{\cal S}^K$ that
corresponds to the $b\rightarrow -\infty$ limit of (\ref{eq:mei}).
Coming from very far apart, the two basic lumps begin to approach
each other when $|b|$ decreases. This approach occurs
simultaneously to a global displacement from $-\infty$ of the
center of mass: $a$ increases. The two kinks merge in a single
TK2E$_-$ lump at the point $b=-1$ in the space ${\cal S}^K$, and
then move together, becoming higher and thinner until they become
the TK1 kink when $b=0$. From this point the composite lump moves
towards the $\vec{\phi}^{B_+}$ kinks, becoming shorter and thicker
until the critical value $b=1$ is reached, where a meiosis of a
TK2E$_+$ kink takes place, giving back rise to two separate lumps.
The geodesic evolution is completed through the increasing
separation of the two lumps, $|b|$ increasing , and the center of
mass motion, $a$ increasing , asymptotically running towards the
$b\rightarrow\infty$ limit of the (\ref{eq:mei}) configuration.

\noindent
\begin{figure}[htbp]
\centerline{ \epsfig{file=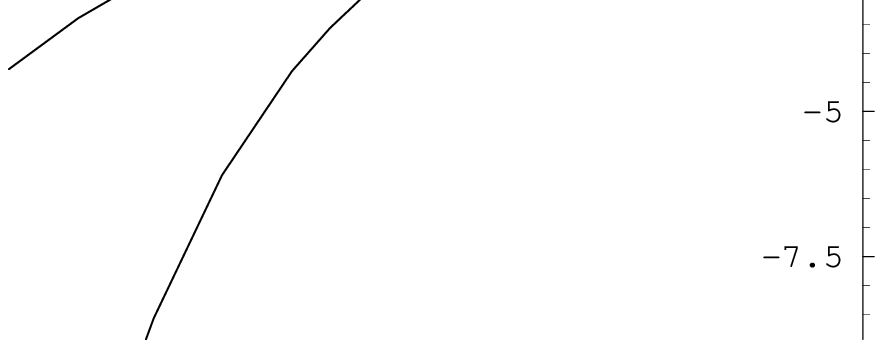, height=3.5cm}\qquad
\epsfig{file=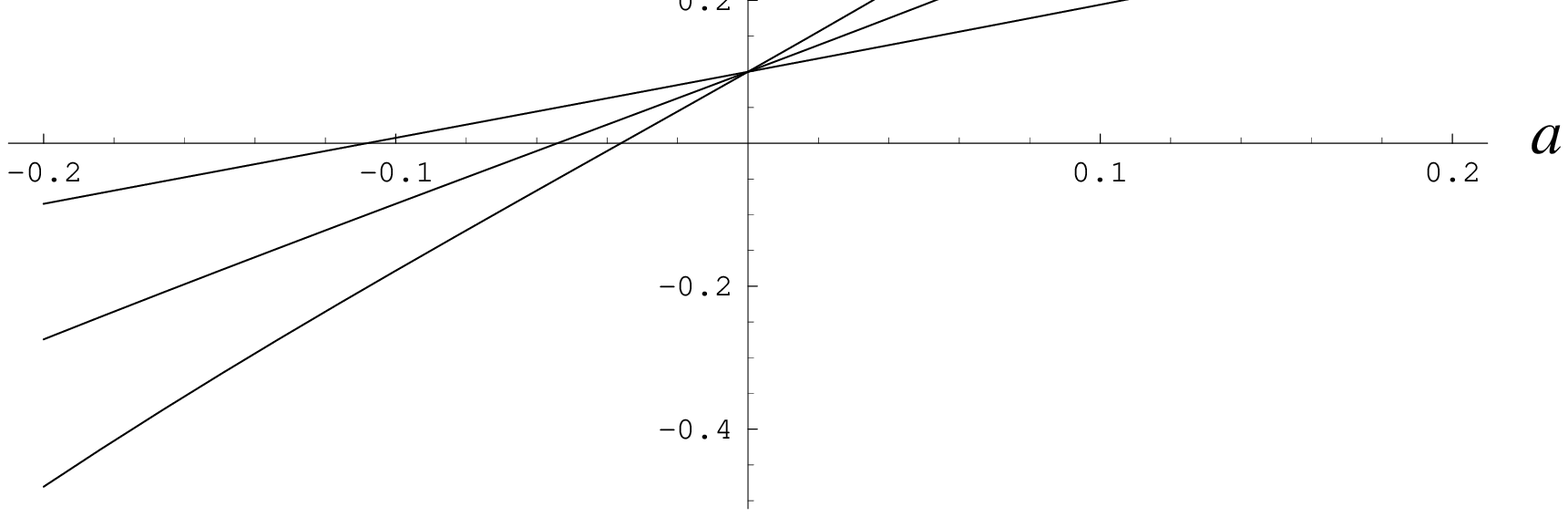, height=3.5cm}} \caption{\small Geodesic
orbits in the $a-b$ plane {\it (a)}.  Detail of the geodesic
orbits {\it (b)}.}
\end{figure}

A closer look at the geodesic orbits near the TK1 point depicted
in Figure 8(b) shows us that the bigger $\kappa_1$ the shorter the
time that the two lumps remain aggregated. At the
$\kappa_1=\infty$ limit, the geodesic curve becomes a vertical
straight line in the space ${\cal S}^K$ and there is no motion of
the center of mass at all. Different values of $\kappa_2$ give
different geodesic orbits by setting the kink point crossed at the
instant $t=-\frac{k_2}{k_1}$.

In summ, a generic orbit describes the following low energy
dynamics: the motion starts from two infinitely separated
$\vec{\phi}^{B_-}[x;x_0,\alpha , \beta]$ kinks (\ref{eq:tkab})
living in the topological sectors in which any configuration
asymptotically connects the vacuum
$\vec{v}^A_-=-\frac{1}{2}\vec{e}_1$ with the vacuum
$\vec{v}^B_+=\vec{e}_2$, and, $\vec{v}^B_+$ with
$\vec{v}^A_+=\frac{1}{2}\vec{e}_1$. The end point, however,
corresponds to two $\vec{\phi}^{B_+}[x;x_0, \alpha ,\beta]$ kinks,
also infinitely separated, (\ref{eq:tkab}) connecting
$\vec{v}^A_-=-\frac{1}{2}\vec{e}_1$ with $\vec{v}^B_-=-\vec{e}_2$,
and, $\vec{v}^B_-$ with $\vec{v}^A_+=\frac{1}{2}\vec{e}_1$. The
whole picture is synthesized in Figure 9. The energy density along
the geodesic $\kappa_1=2$, $k_1=1$, $k_2=0$ is plotted as a
function of $x$ and $t$, showing the adiabatic evolution of the
two basic kinks. In the drawing, time runs from left to right and
the spatial coordinate $x$ grows from bottom to top.

\noindent
\begin{figure}[htbp]
\centerline{ \epsfig{file=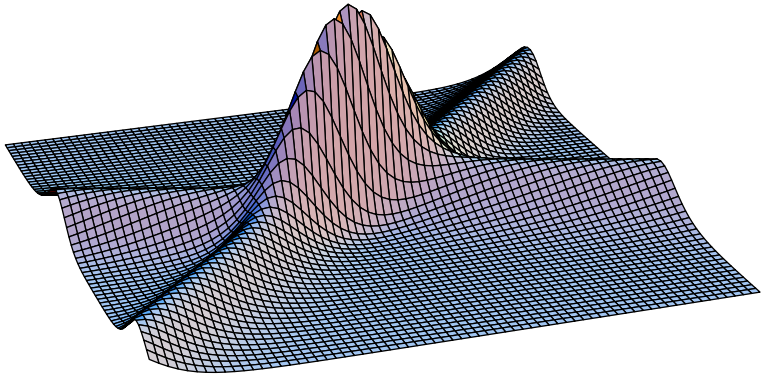, height=5cm}}\caption{\small
Evolution of energy density along  a generic geodesic curve.}
\end{figure}

\subsection{Asymptotic behaviour of geodesics}
It is possible and indeed appropriate to find analytically the
geodesics near the special points $b=0$, $b=\pm\infty$ and $b=\pm
1$ in ${\cal S}^K$ .

\subsubsection{$b \approx 0$}
Close to the TK1 point in ${\cal S}^K$, the metric can be obtained
approximately from the series expansion of $h(b)$ around $b=0$:
\[
h(b)=\, \frac{3\pi}{8}-2b^2+{\cal O}(b^4) \qquad .
\]
The geodesic orbits are given in this region by :
\begin{equation}
\kappa_1a+\kappa_2=\int \sqrt{h(b)} db \simeq \int
\sqrt{\frac{3\pi}{8}-2b^2}\, db ={\frac{4\,b\,{\sqrt{3\,\pi
-16\,{b^2}  }} + 3\,\pi \,\arcsin ({\frac{4\,b}{{ \sqrt{3\,\pi
}}}})}
   {16\,{\sqrt{2}}}} \quad . \label{eq:tk1v}
\end{equation}
Therefore, equation (\ref{eq:tk1v}) analytically describes how the
TK1 kink is reached from kink configurations in its neighbourhood
and vice-versa. Note that this statement is tantamount to saying
that (\ref{eq:tk1v}) determines how the two basic kinks become
completely aggregated on the TK1 kink, $b=-\varepsilon$, and how
they start to split, $b=\varepsilon$.

\subsubsection{$b\approx \pm \infty$}
The expansion of $h(b)$ around the $b=\pm\infty$ points;
\[
h(b)=\frac{1}{2} \, \frac{1}{b^2}+{\cal O}(\frac{1}{b^6})\qquad ,
\]
leads to the geodesic asymptotic behaviour
\begin{equation}
\kappa_1a+\kappa_2=\int \sqrt{h(b)}\, db \simeq
\frac{1}{\sqrt{2}}{\rm sign}(b)\ln |b|\qquad , \label{eq:asy}
\end{equation}
which shows how the two $\vec{\phi}^{B_+}[x;\infty, \alpha ,
\beta]$ kinks (\ref{eq:tkab}) are reached exponentially fast in
$a$, or, how fast the two  $\vec{\phi}^{B_-}[x;-\infty, \alpha ,
\beta]$ lumps start to approach each other.

There is a subtlety: $\lim_{b\rightarrow\pm\infty}h(b)=0$ and the
metric ceases to be a rank-two tensor at these limits. Dynamically
this breakdown of the geometrical meaning is due to the fact that
by taking the $|b|=\infty$ limit we go out of ${\cal S}^K$ because
the separation between the two basic kinks is infinite.
\subsubsection{$b \approx \pm 1$}
There are still two other special points corresponding to the
two-component topological TK2E$_\pm$ kinks where the melting into
a single lump -$b=-1$- or the splitting into two lumps -$b=1$-
take place. Despite appearances, the metric at $b=\pm1$ is regular
and the series expansion of $h(b)$ in the vicinity of these points
reads:
\[
h(b)=\frac{2}{5}\mp\frac{4}{7} (b\mp 1) +{\cal O}((b\mp
1)^2)\qquad .
\]
 Therefore, the geodesic equations
\[
\kappa_1a+\kappa_2=\int \sqrt{h(b)} db \simeq \mp\frac{7}{6}
\left(\frac{34}{35}\mp\frac{4}{7}\, b\right)^{\frac{3}{2}}
\]
analytically rule the low energy process of  two-lump fusion into
TK2E$_-$ / TK2E$_+$ fission into two lumps in the vicinity of the
TK2E$_\pm$ kink points of the ${\cal S}^K$ space .

\section{From low-energy kink dynamics to the slow motion of domain walls}
We finish this paper by offering some comments about the
importance of the model that we have studied within the context of
the low energy effective theories inspired in string/M theory. In
Reference \cite{Voloshin}, the authors analyzed a generalized
${\cal N}=1$ super-symmetric Wess-Zumino model in
$(3+1)$-dimensional space-time with two chiral super-fields ,
$\Phi_1 , \Phi_2$ , and interactions determined by the
super-potential:
\begin{equation}
W(\Phi_1 , \Phi_2
)=\frac{4}{3}\Phi_1^3-\Phi_1+2\sigma\Phi_1\Phi_2^2 \qquad
.\label{eq:supn1}
\end{equation}
We implicitly assume non-dimensional field variables and that
$\sigma$ is a non-dimensional coupling constant between the two
chiral super-fields. Distinguishing among real and imaginary parts
for the super-fields, $\Phi_1=\phi_1+i\psi_1$,
$\Phi_2=\phi_2+i\psi_2$, the real and imaginary parts of the
super-potential $W=W^1+iW^2$  read:
\begin{eqnarray}
W^1(\Phi_1 ,
\Phi_2)&=&\phi_1[\frac{4}{3}\phi_1^2-4\psi_1^2+2\sigma(\phi_2^2-\psi_2^2)-1]
-4\sigma\psi_1\phi_2\psi_2 \label{eq:rsupn1}\\W^2(\Phi_1 ,
\Phi_2)&=&\psi_1[4\phi_1^2-\frac{4}{3}\psi_1^2+2\sigma(\phi_2^2-\psi_2^2)-1]
+4\sigma\phi_1\phi_2\psi_2\nonumber \quad .
\end{eqnarray}
Therefore, the restriction to the real part of the model
-$\psi_1=\psi_2=0 , W^2=0 $ - leads to the BNRT system proposed in
\cite{Bazeia} and discussed from the point of view of kink defects
in \cite{Aai0}. In fact, the BPS kinks described in \cite{Aai0}
are in one-to-one correspondence with the BPS domain walls
discovered in \cite{Voloshin}. In particular, the moduli space of
BPS kinks is identical to the moduli space of BPS walls of the
generalized Wess-Zumino model and all that we have concluded for
the slow motion of BPS kinks in the $\sigma=\frac{1}{2}$ case can
safely be translated to the adiabatic motion of BPS walls in the
corresponding generalized Wess-Zumino model.

This combination of dimensional reduction and reality conditions
poses a problem from a (1+1)-dimensional perspective. We denote by
${\cal W}=W^1$ the real part of the reduced super-potential. It
has been shown in Reference \cite{Voloshin} that there is a
partner \lq\lq quasi-super-potential", $\tilde{\cal W}$, in the
sense that the generalized Cauchy-Riemann equations
\begin{equation}
\frac{\partial{\cal W}}{\partial \phi_a}=2\sigma
\varepsilon_{ab}|\phi_2|^{\frac{2}{\sigma}+1}\frac{\partial\tilde{\cal
W}}{\partial\phi_b}\qquad ;
\varepsilon_{ab}=-\varepsilon_{ba}\quad , a=1,2 \quad ,
b=1,2\qquad .\label{eq:gcr}
\end{equation}
are satisfied . $\mu(\phi_2)=2\sigma
|\phi_2|^{\frac{2}{\sigma}+1}$ is the integrating factor used in
Reference \cite{Aai0} to find the flow lines of ${\rm grad}{\cal
W}$. In fact, the solutions of the first-order equations (9) in
\cite{Aai0} - written here in (\ref{eq:traba1}) for
$\sigma=\frac{1}{2}$ - satisfy:
\[
\frac{\partial{\cal W}}{\partial \phi_2}d\phi_1
-\frac{\partial{\cal W}}{\partial
\phi_1}d\phi_2=\mu(\phi_2)d\tilde{\cal W}=0 \quad .
\]
Thus, $d\tilde{\cal W}$ is an exact one-form on the solutions (in
${\Bbb R}^2$ ) and $\tilde{\cal W}$ remains constant on the kink
orbits. Only if $\sigma =-2$ will the dimensional reduction that
we are considering coincide with the outcome of the standard
dimensional reduction in the genuine Wess-Zumino model. In this
latter case, there is no need for any integrating factor because
(\ref{eq:gcr}) becomes strictly the Cauchy-Riemann equations and
${\cal W},\tilde{\cal W}$ are conjugate harmonic functions that
allow a complex super-potential to be built . If $\sigma\neq -2$
this is not so and there is no possibility of obtaining ${\cal
N}=2$ (1+1)-dimensional super-symmetry, which, in turn, means that
one must expect one-loop corrections in the surface tension of the
walls.

\section*{ACKNOWLEDGEMENT}
We thank to W. Garcia Fuertes for critical reading of the
manuscript.


\begin{thebibliography}{99}

\addcontentsline{toc}{section}{References}

{\small

\bibitem{Manton} N. Manton, Phys. Lett. {\bf B 110} (1982) 52

\bibitem{Uhlenbeck} K.Uhlenbeck, Notices of the AMS {\bf 42}(1995)
41

\bibitem{Gibbons} G.W. Gibbons and P.J. Ruback, Phys. Rev. Lett. {\bf
57} (1986) 1492

\bibitem{Atiyah} M. Atiyah and N. Hitchin, {\sl Geometry and Dynamics
of Magnetic Monopoles}, Princeton University Press, 1988

\bibitem{Samols} T. Samols, Commun. Math. Phys. {\bf 145} (1992) 149

\bibitem{Dziarmaga} J. Dziarmaga, Phys. Rev. {\bf D 51} (1995) 7052

\bibitem{GG} W. Garcia Fuertes and J. Mateos Guilarte, Eur. Phys.
Jour. {\bf C 9} (1999) 535

\bibitem{Manton2} N. Manton, Ann. Phys. {\bf 256} (1997) 114

\bibitem{Kim} Y. Kim and K. Lee, Phys. Rev. {\bf D49} (1994) 2041

\bibitem{Tong} D. Tong, Phys. Rev. {\bf D66} (2002)025013

\bibitem{Townsend} R. Portugues and P.Townsend, Phys. Lett. {\bf
B 530} (2002) 227

\bibitem{Olive}  D. Olive and E. Witten, Phys. Lett. {\bf B 78}
(1987) 97

\bibitem{Bazeia} D. Bazeia , J. Nascimento, R. Ribeiro and D.
Toledo, J. Phys. {\bf A 30} (1997) 8157

\bibitem{Voloshin} M. Shifman and M. Voloshin, Phys. Rev. {\bf
D 57} (1998) 2590

\bibitem{Aai0}  A. Alonso Izquierdo, M.A. Gonzalez Leon and J. Mateos
Guilarte, Phys. Rev. {\bf D 65} (2002) 085012

\bibitem{Sakai} N. Sakai and R. Sugisaka, Phys. Rev. {\bf D66}
(2002) 045010

\bibitem{Coleman} S. Coleman, {\sl Classical lumps and their quantum
descendants}, in {\sl Aspects of Symmetry}, Cambridge University
Press, 1985

\bibitem{Aai1}  A. Alonso Izquierdo, M.A. Gonzalez Leon and J. Mateos
Guilarte, Nonlinearity {\bf 13}  (2000) 1137

\bibitem{Aai2}  A. Alonso Izquierdo, M.A. Gonzalez Leon and J. Mateos
Guilarte,, J. Phys. {\bf A 31} (1998) 209

\bibitem{Aai3} A. Alonso Izquierdo, M.A. Gonzalez Leon and J. Mateos
Guilarte, Phys. Lett. {\bf B 480} (2000) 373



}
\end{thebibliography}
\end{document}